\documentclass[review]{elsarticle}
\graphicspath{ {} }
\usepackage{hyperref}
\usepackage[margin=2.5cm]{geometry}
\usepackage{float}
\usepackage{verbatim} 
\usepackage{apalike}
\usepackage{xcolor}

\usepackage{textcomp, gensymb}
\usepackage{array,ragged2e}
\usepackage[labelformat=empty]{caption}
\newcolumntype{P}[1]{>{\RaggedRight\arraybackslash}p{#1}}
\restylefloat{figure}
\restylefloat{table}

\journal{Expert Systems with Applications}

\biboptions{authoryear}

\begin{document}

\begin{titlepage}
\begin{center}
\vspace*{1cm}

\textbf{ \large Robust and Interpretable COVID-19 Diagnosis on Chest X-ray Images using Adversarial Training}

\vspace{1.5cm}

Karina Yang$^{a}$ (karinaya@usc.edu), Alexis Bennett$^a$ (abbennet@usc.edu), Dominique Duncan$^a$ (duncand@usc.edu) \\

\hspace{10pt}

\begin{flushleft}
\small  
$^a$ Laboratory of Neuro Imaging, USC Stevens Neuroimaging and Informatics Institute, Keck
School of Medicine, University of Southern California, 2025 Zonal Avenue, Los Angeles,
California, USA \\

\vspace{1cm}
\textbf{Corresponding Author:} \\
Karina Yang \\
Laboratory of Neuro Imaging, USC Stevens
Neuroimaging and Informatics Institute, Keck School of Medicine, University of Southern California, 2025 Zonal Avenue, Los Angeles, California, USA
 \\
Tel: (623) 800-5342 \\
Email: karinaya@usc.edu \\
The authors declare no conflicts of interest. 

\end{flushleft}        
\end{center}
\end{titlepage}






\section*{Abstract}
The novel 2019 Coronavirus disease (COVID-19) global pandemic is a defining health crisis. Recent efforts have been increasingly directed towards achieving quick and accurate detection of COVID-19 across symptomatic patients to mitigate the intensity and spread of the disease. Artificial intelligence (AI) algorithms applied to chest X-ray (CXR) images have emerged as promising diagnostic tools, and previous work has demonstrated impressive classification performances. However, such methods have faced criticisms from physicians due to their black-box reasoning process and unpredictable nature. In contrast to professional radiologist diagnosis, AI systems often lack generalizability, explainability, and robustness in the clinical decision making process. In our work, we address these issues by first proposing an extensive baseline study, training and evaluating 21 convolutional neural network (CNN) models on a diverse set of 33,000+ CXR images to classify between healthy, COVID-19, and non-COVID-19 pneumonia CXRs. Our resulting models achieved a 3-way classification accuracy, recall, and precision of up to 97.03\%, 97.97\%, and 99.95\%, respectively. Next, we investigate the effectiveness of adversarial training on model robustness and explainability via Gradient-weighted Class Activation Mapping (Grad-CAM) heatmaps. We find that adversarially trained models not only significantly outperform their standard counterparts on classifying perturbed images, but also yield saliency maps that 1) better specify clinically relevant features, 2) are robust against extraneous artifacts, and 3) agree considerably more with expert radiologist findings.

\textit{Keywords:} Adversarial Training, Computer Vision, COVID-19, Explainable AI, Chest X-rays


\section{Introduction}
\label{introduction}

In recent years, the novel 2019 Coronavirus disease (COVID-19) has posed as a major threat to the lives and well being of people globally. According to the World Health Organization (WHO), there have been over 760 million cases of diagnosed COVID-19, and over 6.8 million deaths attributed to the disease \citep{WHO}. In order to combat these disastrous effects, growing efforts have been directed towards developing early-stage detection mechanisms to reduce the transmission and facilitate the treatment process of the disease.

Currently, the reverse transcription-polymerase chain reaction (RT-PCR) is considered the gold-standard diagnostic tool for COVID-19. In clinical practice, computed tomography (CT) and chest X-ray (CXR) scans have also been popularly employed to assist radiologists in diagnosing COVID-19 with higher certainty \citep{ai2020correlation, kassania2021automatic}. However, this process requires a lot of time and effort, while also remaining susceptible to high levels of human error. Consequently, the application of computer aided diagnosis via artificial intelligence (AI) algorithms has emerged as a promising alternative for more efficient and reliable detection through medical imaging. During peaks of the pandemic, the shortage of radiologists significantly delays the diagnosis and treatment of COVID-19 patients. Thus, a multitude of efforts have been focused on developing and improving automated diagnosis systems to alleviate this problem.

AI algorithms have been proposed as a promising solution, especially given the recent improvements in computational resources and deep convolutional neural network (CNN) based computer vision models. Pre-trained state-of-the-art CNN architectures for COVID-19 diagnosis have seen accuracies up to 99.4\%, though many of these efforts only experiment with a small subset of promising pretrained models \citep{islam2020combined}. Additionally, deep CNN models require large scale datasets to fully demonstrate their discriminative ability. Many works do not meet this demand due to the general lack of publicly accessible CXR data, producing models that may be incapable of generalizing to a different distribution of CXR images. To combat the above concerns, our paper provides a standardized baseline by comparing 21 state-of-the-art computer vision models on a large scale CXR dataset with over 33,000 images, derived from 10 repositories to encourage high predictive power and generalization capabilities.

The classification pipelines of many automated CXR diagnosis systems suffer from a few additional drawbacks. Primarily, deep learning models are susceptible to critical misbehaviors during adversarial attacks, where intentionally perturbed images which are perceived as benign to humans, successfully fool a model into making an incorrect decision with high confidence. In the case of safety-critical settings such as COVID-19 diagnosis, this will lead to high consequence outcomes, such as exacerbating the severity and spread of the disease. Our work addresses this issue by exploring the effect of adversarial training on a subset of the CNN models, creating robust classifiers, which achieve high accuracy on both the standard and adversarially perturbed inputs.

The black-box nature of AI alorithms has long been identified as a major obstacle for usage in clinical settings. To provide explainability of computer vision models, Gradient-weighted Class Activation Mapping (Grad-CAM) is used to provide insight into the model predictions by highlighting important regions in an image. Though recent works have applied this method for transparency in COVID-19 identification in CXRs, the visual explanations are often unclear and susceptible to posit high saliency on irrelevant concepts. For instance, Grad-CAMs may focus on noise beyond the lung tissue region, such as external annotations or medical devices, producing visual explanations that do not clinically pertain to the classified disease \citep{hemied2022covid, lin2020covid, panwar2020deep}. 

Previous works have investigated the influence of adversarial training on the interpretability of CNN models, showing significantly sharper Grad-CAMs than those of standardly trained CNNs. However, to the best of our knowledge, no research has studied the impact of resultant heatmaps from robust models on lung images. Thus, we present the first study that applies adversarial training to improve model explainability in both the lung imaging and COVID-19 diagnosis setting. We find that robust models produce Grad-CAMs that are sharper, more so contained within the lung area, and are also much less susceptible to noisy external artifacts commonly associated with COVID-19 patients.

In summary, our contributions are threefold: 1) we compare the performance of 21 state-of-the-art computer vision models on a large scale CXR dataset derived from 10 repositories to encourage high predictive power and generalization, 2) we demonstrate the strength and importance of robust models in COVID-19 diagnosis, and 3) we show that adversarial training significantly strengthens the visual coherence and interpretability of Grad-CAMs.
 
\section{Related Works}
\label{Related-Works}
\subsection{Computer Vision for COVID-19 Diagnosis}
Many studies have successfully coupled deep learning methods with CXR imaging to diagnose COVID-19. Particularly, transfer learning from pre-trained state-of-the-art image recognition models has appeared as a promising strategy \citep{garg2022efficient}. For example, Vaid et al. used the VGG-19 architecture to classify COVID-19 vs. normal CXR images with 96.3\% accuracy \citep{vaid2020deep}. Minaee et al. and Ismael et al. employed ResNet18, ResNet50, ResNet101, VGG16, and VGG19, SqueezeNet, and DenseNet-121 on CXR images to identify COVID-19 vs. normal CXR images, achieving an accuracy of 90\% and 92.6\%, respectively \citep{ismael2021deep, minaee2020deep}. Similarly, Hemdan et al. applied seven pretrained image classifiers, which were trained and evaluated on 50 X-ray images (25 normal and 25 COVID-19 cases), obtaining 90.00\% accuracy \citep{hemdan2020covidx}. Marques et al. applied EfficientNet architectures to a CNN to conduct binary classification (COVID-19 vs. normal), and multi-classification (COVID-19 vs. pneumonia vs. normal) achieving 99.6\% and 96.7\% accuracy, respectively \citep{marques2020automated}. Wang et al. trained a deep CNN to perform a 3-way classification of patients with COVID-19, pneumonia, and no infection, reporting 93.3\% accuracy. Heidari et al., Apostolopoulos et al., and Boudrioua likewise exploited a multiclass transfer learning approach to classify between COVID-19, non-COVID-19 pneumonia, and normal cases, attaining up to 94.5\% and 96.7\% accuracy, and 99.5\% sensitivity rates, respectively \citep{apostolopoulos2020covid, boudrioua2020covid, heidari2020improving}. Khan et al. presented CoroNet, applying the pre-trained Xception, to perform 3 and 4-way multi-classifications, reporting 89.6\% and 95.0\% accuracy, respectively, whereas Keles et al. proposed distinguishing between COVID-19, viral pneumonia, and normal CXR images with their COV19-ResNet model, achieving a high accuracy of 97.6\% \citep{keles2021cov19, khan2020coronet}. 

Recent efforts have also focused on developing novel deep learning architectures and pipelines for COVID-19 identification. Hussain et al. developed a 22-layer CNN-based model named CoroDet for 2-way classification (COVID-19 vs. normal) and multi-classifications, reporting accuracy rates up to 94.2\% \citep{hussain2021corodet}. Ozturk et al. proposed DarkCovidNet, a deep learning network that obtained an accuracy of 87.02\% for multi-class classification (COVID-19 vs. normal, vs. non-COVID-19 pneumonia), and 98.08\% for binary classification (COVID-19 vs. normal) on CXR images \citep{ozturk2020automated}. A shallow CNN was designed by Mukherjee et al. to differentiate between COVID-19 and non-COVID-19 classes with 99.69\% accuracy \citep{mukherjee2021shallow}. Ezzat et al. used gravitational search optimization-DenseNet121-Covid-19 to classify between COVID-19 and non-COVID-19 CXRs with 93.4\% accuracy, while Babukarthik et al. applied genetic deep CNN to detect COVID-19 from two classes of CXR images, reporting high performance accuracy (98.8\%) \citep{babukarthik2020prediction, ezzat2021optimized}.  Islam et. al combined a CNN feature extractor and a long short-term memory (LSTM) classifier to differentiate between patients with COVID-19, non-COVID-19 pneumonia, and no infection, with 99.4\% accuracy \citep{islam2020combined}. Karthik et al. employed a channel shuffled dual-branched residual CNN to perform a 4-way classification between COVID-19, bacterial pneumonia, viral pneumonia, and normal class images, achieving 99.8\% accuracy \citep{karthik2021learning}. 

Relevant deep learning for COVID-19 diagnosis using CXR images are reviewed and summarized in Table 1. 

\begin{table}[H]
\caption{Table 1: Summary of Recent Deep Learning Efforts for COVID-19 Classification from Chest X-rays}
\resizebox{\columnwidth}{!}{\begin{tabular}{P{3cm} P{5.5cm} P{8cm} P{1.3cm} P{3.3cm}}
\hline
\textbf{References} & \textbf{Model(s)} & \textbf{Number of Images} & \textbf{Classes} & \textbf{Highest Reported Accuracy} \\ 
\hline 
Wang et al.
& COVID-Net
& 1125 - total; 125 - COVID-19; 500 - Pneumonia; 500 - Normal
& 3
& 93.3\% \\

Hemdan et al. 
& COVIDXNet (based on VGG19, DenseNet201, InceptionV3, ResNetV2, InceptionResNetV2, Xception, MobileNetV2)
& 50 - total; 25 - COVID-19; 25 - Normal 
& 2
& 90.0\% \\

Vaid et al.
& VGG-19
& 545 - total; 181 - COVID-19; 364 - Normal
& 2
& 96.3\% \\

Heidari et al.
& VGG-16-based CNN
& 8474 - total; 415 - COVID-19; 5179 - Pneumonia; 2880 - Normal;
& 3
& 94.5\% \\

Apostolopoulos et al.
& VGG19 and MobileNet v2
& Dataset 1: 1427 - total; 224 - COVID-19; 700 - Pneumonia; 504 - Normal; Dataset 2: 1442 - total; 224 - COVID-19; 714 - Pneumonia; 504 - Normal; 
& 3
& 96.7\% \\

Ismael et al.
& ResNet18, ResNet50, ResNet101, VGG16, VGG19
& 380 - total; 180 - COVID-19; 200 - Normal 
& 2
& 92.6\% \\

Islam et al.
& CNN + LSTM
& 4575 - total; 1525 - COVID-19; 1525 - Pneumonia; 1525 - Normal; 
& 3
& 99.4\% \\

Karthik et al.
& Dual-branched residual CNN
& 15,265 - total; 558 - COVID-19; 2780 - Bacterial Pneumonia; 1493 - Viral Pneumonia; 10,434 - Normal;
& 4
& 97.9\% \\

Khan et al.
& CoroNet
& 1251 - total; 284 - COVID-19; 330 - Bacterial Pneumonia; 327 - Viral Pneumonia; 310 - Normal;
& 3, 4
& 95.0\%, 89.6\% \\

Oh et al.
& ResNet-18
& 502 - total; 180 - COVID-19; 54 - Bacterial Pneumonia; 20 - Viral Pneumonia; Tuberculosis - 57; 191 - Normal;
& 5
& 91.9\% \\

Ozturk et al.
& DarkNet
& 1127 - total; 127 - COVID-19; 500 - Pneumonia; 500 - Normal; 
& 2, 3
& 98.1\%, 87.0\% \\

Minaee et al.
& ResNet18, ResNet50, SqueezeNet, DenseNet-121
& 5184 - total; 184 - COVID-19; 5000 - Normal 
& 2
& 90.0\% \\

Boudrioua
& DenseNet 121, NASNetLarge and NASNetMobile
& 3309 - total; 309 - COVID-19; 2000 - Pneumonia; 1000 - Normal;
& 3
& 99.5\% (sensitivity) \\

Ezzat et al.
& Gravitational search optimization (GSA) -DenseNet121-COVID-19
& 306 - total; 99 - COVID-19; 207 - Other
& 2 
& 93.4\% \\

Marques et al.
& CNN + EfficientNet
& 1508 - total; 504 - COVID-19; 504 - Pneumonia; 500 - Normal; 
& 2, 3
& 99.6\%, 96.7\% \\

Babukarthik et al. 
& Genetic deep CNN
& 5536 - total; 536 - COVID-19; 5000 - Normal
& 2
& 98.8\% \\

Hussain et al.
& CoroDet
& 2100 - total; 500 - COVID-19; 400 - Bacterial Pneumonia; 400 - Viral Pneumonia; 800 - Normal;
& 2, 3, 4
& 99.1\%, 94.2\%, 91.2\% \\

Mukherjee et al.
& Shallow CNN
& 4594 - total; 321 - COVID-19; 4273 - Pneumonia; 
& 2
& 99.7\% \\

Keles et al.
& COV19-ResNet
& 910 - total; 210 - COVID-19; 350 - Pneumonia; 350 - Normal;
& 3
& 97.6\% \\
\hline

\end{tabular}}
\end{table}

Similar to the works of \cite{wang2020covid}, \cite{showkat2022efficacy}, \cite{chakraborty2022transfer}, \cite{ozturk2020automated}, \cite{boudrioua2020covid}, \cite{marques2020automated}, \cite{hussain2021corodet}, \cite{keles2021cov19}, etc., our study classifies between COVID-19, Pneumonia, and Normal Chest X-ray scans. Of these recent efforts, \cite{hemdan2020covidx} applies transfer learning to the largest number of state of the art deep learning models (7), whereas we obtain performance results for 21 different models. In Table 1, the highest reported accuracy for 3-way classification using transfer learning is achieved by \cite{keles2021cov19} (97.6\%), which is comparable to our highest attained accuracy of 97.03\%. Out of all the related works, our data set used for training, validating, and testing the models is the largest, with a total of 33,920 samples from 10 different repositories. Given that our train and test set are both patient and repository independent, we achieve a satisfiable performance measured by accuracy.

\subsection{Gradient-based Visual Explanations}
In addition to strong classification performance of deep learning models, recent efforts have also emphasized the importance of explainability in AI algorithms by providing insight into the model’s decision making process. Selvaraju et al. proposed Grad-CAM, which produces a coarse localization map that highlights important regions of the image when making a prediction \citep{selvaraju2017grad}. GradCAMs have been popularly applied to computer vision based COVID-19 diagnosis \citep{basu2020deep, karim2020deepcovidexplainer, oh2020deep, panwar2020deep}; however, the visual indicators are periodically inconsistent with radiologist findings, or are heavily influenced by clinically irrelevant external artifacts or annotations. Since the introduction of adversarial training by Szegedy et al, current research efforts have combined this technique with Grad-CAM visualizations for more consistent and distinct model visualizations \citep{szegedy2013intriguing}. For instance, Zhang et al. used adversarial training to improve interpretability of deep learning networks by demonstrating more visually coherent SmoothGrad saliency maps, and the alleviation of texture bias on classifying natural images \citep{zhang2019interpreting}. Margeloiu et al. extended this work by investigating the influence of adversarial training on CNN-based skin cancer diagnosis, demonstrating sharper and more visually coherent saliency maps (integrated gradients, occlusion, Grad-CAMs, etc.) \citep{margeloiu2020improving}. We apply methods used by \cite{zhang2019interpreting} and \cite{margeloiu2020improving} in the COVID-19 diagnosis setting on multiple state-of-the-art computer vision models. We analyze the improvement in visual coherence and explainabilty on lung X-ray images, as opposed to natural images used by \cite{zhang2019interpreting} and skin cancer images used by \cite{margeloiu2020improving}.

Most previous work has demonstrated the significance of deep learning for the detection of COVID-19 through patient CXR images. These studies have either explored variations of state-of-the-art image classification models, or proposed modifications to existing deep learning architectures. The table and this section, however, highlight the lack of consensus for base model usage. Additionally, existing literature does not consider the adversarial vulnerabilities of proposed models, nor the inconsistent explainability power of vanilla Grad-CAMs. The proposed methods in this paper make several novel contributions to existing literature. We first provide a consistent backbone for future efforts by comparing the performance of 21 state of the art image classification models on a large dataset of over 33,000 CXR images. We then demonstrate how adversarially trained models exhibit robust performance against perturbed images. Finally, we utilize adversarial training and Grad-CAMs to reinforce the interpretability of black-box deep learning models in the COVID-19 diagnosis setting. 

\section{Methods}
\label{Methods}
\subsection{Dataset}
This work used the COVID-QU-Ex dataset, which was comprised of 33,920 CXR images, with 11,956 COVID-19, 11,263 non-COVID-19 infections (Viral or Bacterial Pneumonia), 10,701 normal images from 10 different CXR databases \citep{covidquex}. The source dataset provides Lung Segmentation Masks for all CXR images, and infection masks for 2,913 COVID-19 CXRs. All images are grayscale, and freely accessible through Kaggle. The authors of this database have additionally divided all images into the training, validation, and test set, where images between the different sets are both patient and repository independent. Quantitative details are presented in Table 2.

\begin{table}[H]
\caption{Table 2: COVID-QU-Ex Dataset Quantitative Information}
\resizebox{\columnwidth}{!}{\begin{tabular}{P{3cm} P{3cm} P{3cm} P{3cm} P{3cm} P{4cm}}
\hline
\textbf{Dataset} & \textbf{COVID-19} & \textbf{Non-COVID-19 Pneumonia} & \textbf{Normal} & \textbf{Total} & \textbf{Number of COVID-19 Infection Masks}\\ 
\hline 
Train & 7,658 & 7,208 & 6,849 & 21,715 & 1,864 \\
Validation & 1,903 & 1,802 & 1,712 & 5,417 & 466 \\
Test & 2,395 & 2,253 & 2,140 & 6,788 & 583 \\
Total & 11,956 & 11,263 & 10,701 & 33,920 & 2,913 \\
\hline

\end{tabular}}
\end{table}

The dataset can be found on the COVID-19 Data Archive (COVID-ARC) \citep{duncan2021covid} and can be accessed at https://covid-arc.loni.usc.edu/\#dataset.
\subsection{Data Preprocessing}
Data preprocessing was applied for robust and generalizable model classification. To leverage patient-independent features for multi-class classification (COVID-19, non-COVID-19 pneumonia, normal), we utilized the training, validation, and test sets provided by the authors of the dataset. We applied image augmentation to the training set of images to boost the diversity of images within the training data and reduce overfitting of our deep learning models on the test set. Specifically, we used the ImageDataGenerator API in Keras, which generates batches of tensor image data with real-time data augmentations of processing, such as random rotation, shifts, shear and flips, etc. \citep{chollet2015keras}. For each sample image, the pixels were rescaled to the range $[0,1]$, horizontal and vertical flips were randomly applied, and random values (selected within the ranges presented in Table 3) were applied for zoom, rotation, width, height, and shear. The number of image samples remains unchanged. A quantitative summary of applied augmentations are summarized in Table 3 below.

\begin{table}[H]
\centering
\begin{tabular}[t]{cc}
\hline
Pixel Rescaling Factor & 1./255 \\ 
Horizontal Flips Allowed & True \\ 
Vertical Flips Allowed & True \\
Zoom Range & [0.80, 1.20] \\
Rotation Range & [0\degree, 180\degree] \\
Width Shift Range & [-20\%, +20\%] \\
Height Shift Range & [-20\%, +20\%] \\
Shear Range & [-10\%, +10\%] \\
\hline
\end{tabular}
\caption{Table 3: Summary of Applied Data Augmentation Techniques}
\end{table}

\subsection{Model Development and Training}
A total of 21 models were trained and evaluated in this study. Each of these models were derived from the following base models: ResNet 50 \citep{he2016deep}, ResNet 50V2 \citep{he2016identity}, ResNet 101V2 \citep{he2016deep}, ResNet 152V2 \citep{he2016deep}, InceptionV3 \citep{szegedy2016rethinking}, InceptionResNetV2 \citep{szegedy2017inception}, Xception \citep{chollet2017xception}, EfficientNet V2S \citep{tan2021efficientnetv2}, EfficientNet V2M \citep{tan2021efficientnetv2}, EfficientNet V2B0 \citep{tan2021efficientnetv2}, EfficientNet V2B1 \citep{tan2021efficientnetv2}, EfficientNet V2B2 \citep{tan2021efficientnetv2}, EfficientNet V2B3 \citep{tan2021efficientnetv2}, EfficientNet B4 \citep{tan2019efficientnet}, EfficientNet B5 \citep{tan2019efficientnet}, EfficientNet B6 \citep{tan2019efficientnet}, DenseNet 121 \citep{huang2017multi}, DenseNet 169 \citep{huang2017multi}, DenseNet 201 \citep{huang2017multi}, VGG16 \citep{simonyan2014very}, and VGG19 \citep{simonyan2014very}. These models were chosen due to their exemplary performance on the ImageNet dataset and their promising potential demonstrated in section \ref{Related-Works}.

For our experiments, we maintain the architecture of the above models, initializing the weights with the ImageNet versions. The final classification layer of the models were adapted from 1000-dimensional to 3 dimensional nodes, pertaining to the 3 classes in our dataset: normal, COVID-19, and non-COVID-19 pneumonia. All model weights were unfrozen, and models were trained on 4 V100-SXM2-32GB GPUs with distributed training on TensorFlow 2.10.0. Given our sufficiently large data size and computational power, we hope that retraining the weights of pretrained model architectures yields stronger classification power than any variation of freezing and fine-tuning.

We utilized the Adaptive Moment Estimation (‘adam’) Optimizer with a learning rate of 0.0001, and applied a reduced learning rate on plateau callback, which decreased the optimizer learning rate by a factor of 0.2 after 2 consecutive epochs where the decreases in validation loss were 0.0001 or less. We trained our model using a batch size of 32 for 100 maximum epochs, and terminated training once performance demonstrated no improvement over 15 epochs in order to prevent overfitting. For all models, the best performing model state across all epochs was maintained and used to calculate final performance metrics on the testing dataset. For model validation, this process was repeated for a total of 15 rounds of training and evaluation. The average value of metrics across all rounds was computed and their expected values are presented within a 95\% confidence interval.

\subsection{Robust Model Development and Training}
This work also investigates the performance and explainability of deep learning models on adversarially perturbed examples. Adversarial images are intentionally perturbed for inducing incorrect model classification, though are generally indistinguishable to the human eye. Figure 1 depicts both imperceptible and perceptible pixel-level distortions appended to a COVID-19 positive CXR, resulting in the VGG16 model labeling the image incorrectly.

\begin{figure}[H]
\centering
\caption{Figure 1: Demonstration of Adversarial Perturbations on COVID-19 CXR}
\includegraphics[scale = 0.8]
{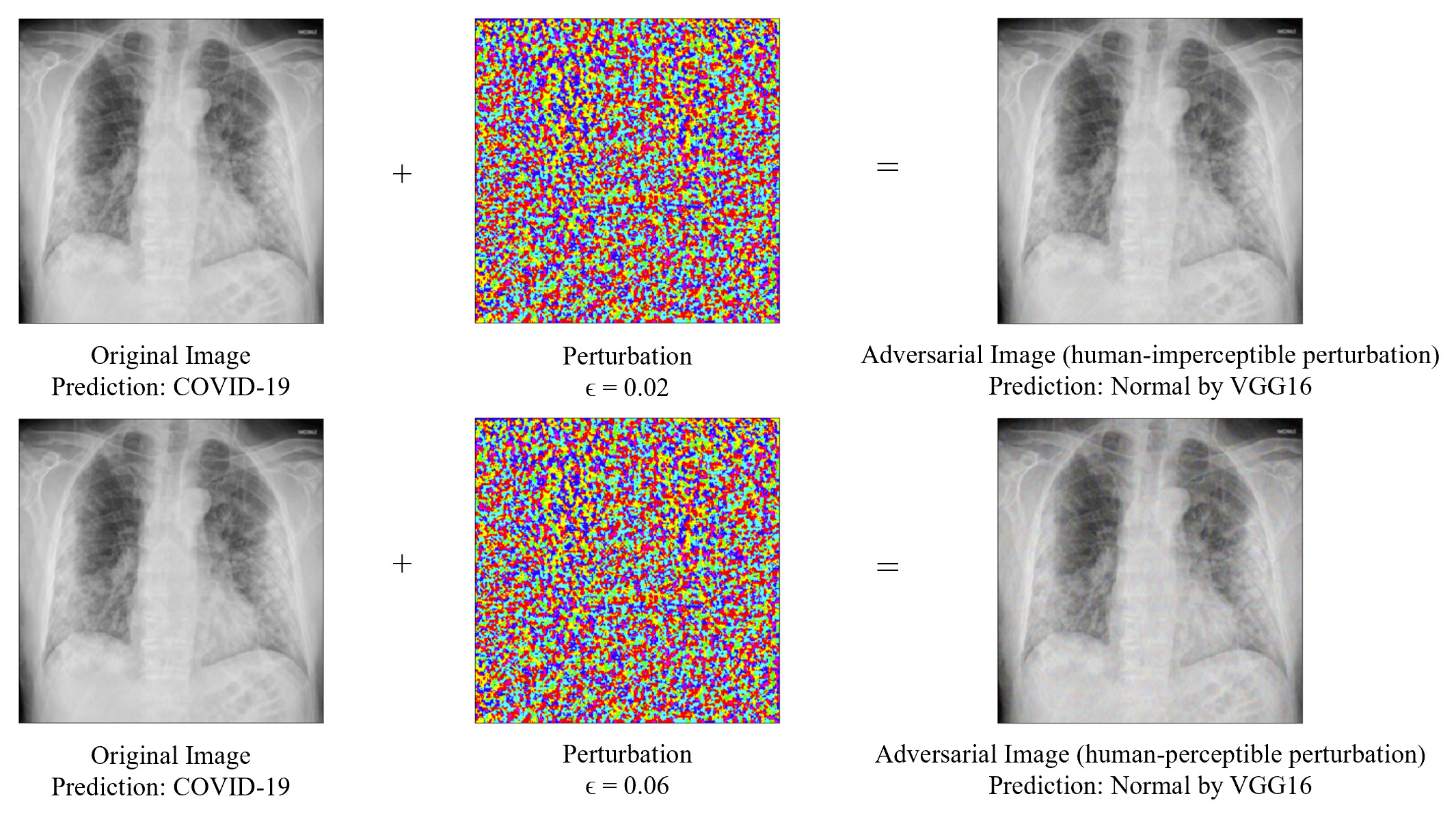}
\end{figure}

Robust classifiers are models that achieve high accuracy on adversarially perturbed inputs \citep{szegedy2013intriguing}. The common method to obtain robust models is to apply adversarial training, where, at each training iteration, imperceptible pixel-level perturbations are dynamically added in the reverse gradient direction with respect to the input.

Specifically, we apply the fast gradient sign method for creating perturbed images \citep{goodfellow2014explaining}, as done in a few previous works \citep{zhang2019interpreting, margeloiu2020improving}. Let $\theta =$ the model parameters, $x =$ the model input, $y =$ the target classes associated with $x$, and $J(\theta, x, y) =$ the cost used to train the model. The cost function, $J$, is linearized around the current value of $\theta$, obtaining an optimal max-norm constrained perturbation of $\eta = \epsilon$\textit{sign}$(\nabla_x J(\theta, x, y))$. Varying the value of $\epsilon$, usually from 0 to 1, produces a range of adversarial examples that may be more or less perceptible to the human eye.

Previous work on COVID-19 classification has demonstrated that as the $\epsilon$ value increases, the percentage of incorrect predictions increases \citep{pal2021vulnerability}. In our experiments, we choose to apply $\epsilon = 0.02$, the highest $\epsilon$ value yielding an adversarial image that remains generally imperceptible for the human eye \citep{pal2021vulnerability}. We believe this critical value is most suitable for experimentation since it creates input images that, while seemingly unperturbed to the human eye, cause the lowest performance accuracy. Any $\epsilon > 0.02$ would be easily detectable by the human eye, decreasing the likelihood of misclassification going unnoticed. Any $\epsilon < 0.02$ would cause a smaller performance drop of standard models validated on perturbed inputs \citep{pal2021vulnerability}. Thus, we utilize $\epsilon = 0.02$ as a critical value for the experiments.

Adversarial training is applied to a subset of the 21 tested models with identical hyper-parameters compared to standard models. In this way, the model architecture and training procedures do not require any additional adjustments. To select a subset of our 21 tested models for adversarial training experiments, we take the top performing model from each family of models, totalling 6, to conduct performance comparisons between standard to adversarially trained models with identical metrics. The six selected models were EfficientNetV2S, InceptionV3, Xception, ResNet50V2, VGG19, and DenseNet201. We similarly ran 15 rounds of training and evaluation, and report the final performance on the testing data.

\subsection{Visualizations}
Our work further compares the visualization quality of generated Grad-CAMs \citep{selvaraju2017grad} between standard and adversarially trained models. Grad-CAMs are effective in unraveling the black-box nature of decision-making within deep learning algorithms by highlighting the relative important regions of an image for making a classification. By offering human-interpretable insights, they offer an explainable avenue for why a particular decision was made, and increase the chances of being viewed favorably by human evaluators.

We apply Grad-CAMs to the 6 models that were both normally and adversarially trained (EfficientNetV2S, InceptionV3, Xception, ResNet50V2, VGG19, and DenseNet201) to offer a visual comparison. Following previous work in model interpretability, we make qualitative comparisons between the visual coherence of robust and standard models \citep{margeloiu2020improving, smilkov2017smoothgrad, sundararajan2017axiomatic} on standard input images. In our experiments, visual coherence means that salient areas primarily highlight regions of interest (areas contained within the lung). We conduct comparisons of Grad-CAM heatmaps for combinations of different models and training procedures on non perturbed input images of test set COVID-19 CXRs. All presented heatmaps are generated from models that correctly classified the CXRs as COVID-19.

\section{Results}
\subsection{Model Performances}
Table 4 summarizes the performance of standard models on COVID-19 classification in the unseen test set using the metrics: accuracy, precision, recall, and F1 score. EfficientNetV2S obtained the highest accuracy (0.9703), recall (0.9797), and F1 score (0.9860), while Xception achieved the highest precision value (0.9995). Highest performances for each metric are bolded and underlined.

\begin{table}[H]
\caption{Table 4: Summary of Standard Model Performances for COVID-19 Classification}
\resizebox{\columnwidth}{!}{\begin{tabular}{P{0.5cm} P{3cm} P{3cm} P{3cm} P{3cm} P{3cm}}
\hline
\textbf{\#} & \textbf{Model} & \textbf{Accuracy} & \textbf{Precision
} & \textbf{Recall} & \textbf{F1-score}\\ 
\hline 
1 & ResNet50 & 0.9209 $\pm$ 0.0058 & 0.9828 $\pm$ 0.0067 & 0.9011 $\pm$ 0.0063 & 0.9402 $\pm$ 0.0052 \\
2 & ResNet50V2 & 0.9423 $\pm$ 0.0098 & 0.9701 $\pm$ 0.0099
 & 0.9578 $\pm$ 0.0120 & 0.9639 $\pm$ 0.0106 \\
3 & ResNet152V2 & 0.9406 $\pm$ 0.0070 & 0.9632 $\pm$ 0.0072 & 0.9482 $\pm$ 0.0075 & 0.9556 $\pm$ 0.0088 \\
4 & ResNet101V2 & 0.9370 $\pm$ 0.0108 & 0.9742 $\pm$ 0.0116 & 0.9782 $\pm$ 0.0114 & 0.9762 $\pm$ 0.0096 \\
5 & EfficientNetV2S & \underline{\textbf{0.9703 $\pm$ 0.0065}} & 0.9923 $\pm$ 0.0062 & \underline{\textbf{0.9797 $\pm$ 0.0060}} & \underline{\textbf{0.9860 $\pm$ 0.0059}} \\
6 & EfficientNetV2M & 0.9313 $\pm$ 0.0134 & 0.9756 $\pm$ 0.0153
 & 0.9258 $\pm$ 0.0122 & 0.9500 $\pm$ 0.0143 \\
7 & EfficientNetV2B0 & 0.8974 $\pm$ 0.0078 & 0.9196 $\pm$ 0.0071
& 0.9383 $\pm$ 0.0068 & 0.9289 $\pm$ 0.0082 \\
8 & EfficientNetV2B1 & 0.8980 $\pm$ 0.0051 & 0.9537 $\pm$ 0.0059 & 0.8760 $\pm$ 0.0052 & 0.9132 $\pm$ 0.0058 \\
9 & EfficientNetV2B2 & 0.9200 $\pm$ 0.0054 & 0.9517 $\pm$ 0.0059
& 0.9352 $\pm$ 0.0055 & 0.9434 $\pm$ 0.0060 \\
10 & EfficientNetV2B3 & 0.9309 $\pm$ 0.0068 & 0.9678 $\pm$ 0.0062 & 0.9142 $\pm$ 0.0108 & 0.9402 $\pm$ 0.0077 \\
11 & EfficientNetB4 & 0.8959 $\pm$ 0.0102 & 0.9734 $\pm$ 0.0088 & 0.8829 $\pm$ 0.0092 & 0.9259 $\pm$ 0.0105 \\
12 & EfficientNetB5 & 0.9436 $\pm$ 0.0063 & 0.9670 $\pm$ 0.0065 & 0.9333 $\pm$ 0.0064 & 0.9499 $\pm$ 0.0064 \\
13 & EfficientNetB6 & 0.8903 $\pm$ 0.0047 & 0.9882 $\pm$ 0.0088 & 0.8504 $\pm$ 0.0051 & 0.9188 $\pm$ 0.0055 \\
14 & VGG16 & 0.9271 $\pm$ 0.0111 & 0.9849 $\pm$ 0.0160 & 0.9582 $\pm$ 0.0130 & 0.9714 $\pm$ 0.0119 \\
15 & VGG19 & 0.9438 $\pm$ 0.0159 & 0.9612 $\pm$ 0.0102 & 0.9589 $\pm$ 0.0098 & 0.9600 $\pm$ 0.0108 \\
16 & DenseNet121 & 0.9209 $\pm$ 0.0064 & 0.9564 $\pm$ 0.0058 & 0.9410 $\pm$ 0.0097 & 0.9486 $\pm$ 0.0067 \\
17 & DenseNet169 & 0.9366 $\pm$ 0.0053 & 0.9812 $\pm$ 0.0069 & 0.9292 $\pm$ 0.0121 & 0.9545 $\pm$ 0.0061 \\
18 & DenseNet201 & 0.9528 $\pm$ 0.0070 & 0.9795 $\pm$ 0.0061 & 0.9460 $\pm$ 0.0080 & 0.9625 $\pm$ 0.0068 \\
19 & InceptionResNetV2 & 0.9614 $\pm$ 0.0069 & 0.9910 $\pm$ 0.0076 & 0.9675 $\pm$ 0.0065 & 0.9791 $\pm$ 0.0073 \\
20 & InceptionV3 & 0.9389 $\pm$ 0.0070 & 0.9711 $\pm$ 0.0073 & 0.9260 $\pm$ 0.0059 & 0.948 $\pm$ 0.0074 \\
21 & Xception & 0.9518 $\pm$ 0.0072 & \underline{\textbf{0.9995 $\pm$ 0.0060}} & 0.9505 $\pm$ 0.0067 & 0.9744 $\pm$ 0.0081 \\

\hline

\end{tabular}}
\end{table}

Tables 5 and 6 summarize the classification capabilities of standard and robust models on non-perturbed and perturbed images of COVID-19 CXRs. Classification performances of models evaluated on perturbed images are denoted with an '\textbf{*}'. Highest performances of models evaluated on standard and perturbed images for each metric are bolded and underlined, respectively. 

\begin{table}[H]
\caption{Table 5: Summary of Standard Model Performances for COVID-19 Classification on Perturbed Images}
\resizebox{\columnwidth}{!}{\begin{tabular}{ P{3cm} P{3cm} P{3cm} P{3cm} P{3cm}}
\hline
\textbf{Model} & \textbf{Accuracy} & \textbf{Precision
} & \textbf{Recall} & \textbf{F1-score}\\ 
\hline 
EfficientNetV2S\textbf{*} & 0.6459 $\pm$ 0.0157 & \underline{0.8281 $\pm$ 0.0142}
 & \underline{0.7623 $\pm$ 0.0168} & \underline{0.7938 $\pm$ 0.0165} \\
InceptionV3\textbf{*} & 0.4572 ± 0.0101 & 0.5494 ± 0.0094 & 0.5002 ± 0.0087 & 0.5236 ± 0.0090 \\
Xception\textbf{*} & 0.4701 ± 0.0171 & 0.6230 ± 0.0157 & 0.5261 ± 0.0159 & 0.5705 ± 0.0152 \\
ResNet50V2\textbf{*} & \underline{0.6895 ± 0.0069} & 0.7260 ± 0.0073 & 0.7141 ± 0.0066 & 0.7200 ± 0.0074 \\
VGG19\textbf{*} & 0.5111 ± 0.0128 & 0.6656 ± 0.0131 & 0.5827 ± 0.0148 & 0.6214 ± 0.0105 \\
DenseNet201\textbf{*} & 0.6120 ± 0.0143 & 0.5685 ± 0.0132
 & 0.5400 ± 0.0139 & 0.5539 ± 0.0144\\
\hline

\end{tabular}}
\end{table}

\begin{table}[H]
\caption{Table 6: Summary of Robust Model Performances for COVID-19 Classification on Standard and Perturbed Images}
\resizebox{\columnwidth}{!}{\begin{tabular}{ P{3cm} P{3cm} P{3cm} P{3cm} P{3cm}}
\hline
\textbf{Model} & \textbf{Accuracy} & \textbf{Precision
} & \textbf{Recall} & \textbf{F1-score}\\ 
\hline 
EfficientNetV2S & \textbf{0.9621 ± 0.0083
} & 0.9830 ± 0.0092 & \textbf{0.9874 ± 0.0071} & \textbf{0.9852 ± 0.0077
} \\
EfficientNetV2S\textbf{*} & \underline{0.9503 ± 0.0114} & 0.9690 ± 0.0121
 & 0.9649 ± 0.0126 & 0.9669 ± 0.0110 \\
InceptionV3 & 0.9391 ± 0.0082 & 0.9615 ± 0.0097 & 0.9660 ± 0.0058 & 0.9637 ± 0.0056 \\
InceptionV3\textbf{*} & 0.9217 ± 0.0088 & 0.9487 ± 0.0091 & 0.9761 ± 0.0095 & 0.9622 ± 0.0102 \\
Xception & 0.9442 ± 0.0095 & 0.9470 ± 0.0072 & 0.9533 ± 0.0087 & 0.9501 ± 0.0078 \\
Xception\textbf{*} & 0.9401 ± 0.0109 & \underline{0.9923 ± 0.0104} & 0.9499 ± 0.0096
& 0.9706 ± 0.0097 \\
ResNet50V2 & 0.9528 ± 0.0094 & 0.9761 ± 0.0097 & 0.9642 ± 0.0092 & 
0.9701 ± 0.0105 \\
ResNet50V2\textbf{*} & 0.9313 ± 0.0126 & 0.9629 ± 0.0111 & 0.9537 ± 0.0088 & 
0.9583 ± 0.0096 \\
VGG19 & 0.9244 ± 0.0063 & 0.9218 ± 0.0069 & 0.9305 ± 0.0078 & 0.9261 ± 0.0074 \\
VGG19\textbf{*} & 0.9270 ± 0.0096 & \textbf{0.9882 ± 0.0074} & 0.9222 ± 0.0080 & 0.9541 ± 0.0085 \\
DenseNet201 & 0.9618 ± 0.0077 & 0.9804 ± 0.0091 & 0.9639 ± 0.0086 & 
0.9721 ± 0.0076 \\
DenseNet201\textbf{*} & 0.9482 ± 0.0090 & 0.9668 ± 0.0108 & \underline{0.9878 ± 0.0102} & \underline{0.9772 ± 0.0098}\\
\hline

\end{tabular}}
\end{table}

\subsection{Grad-CAM Visualization Comparisons}
All Grad-CAM saliency maps are computed on correctly-classified COVID-19 CXR images and are visualized as RGB-colored heatmaps. The Grad-CAM highlights the regions that positively support predicting the class of interest (COVID-19). 

\subsubsection{General Models}
The figure below depict the Grad-CAM visualization comparison across both the standard and adversarially trained models: EfficientNetV2S, InceptionV3, Xception, ResNet50V2, VGG19, and DenseNet201 for a CXR image of a COVID-19 patient. Regardless of model training procedures, we recognize that Xception, DenseNet201, and VGG19 generally offer more visually coherent heatmaps. Compared to robust models, standard models present heatmaps that are more sparse, and often focus on external medical devices (top left), annotations (bottom left), and regions outside of the lung.

\begin{figure}[H]
\centering
\caption{Figure 2: Standard and Robust Model Grad-CAMs}
\includegraphics[scale = 0.8]
{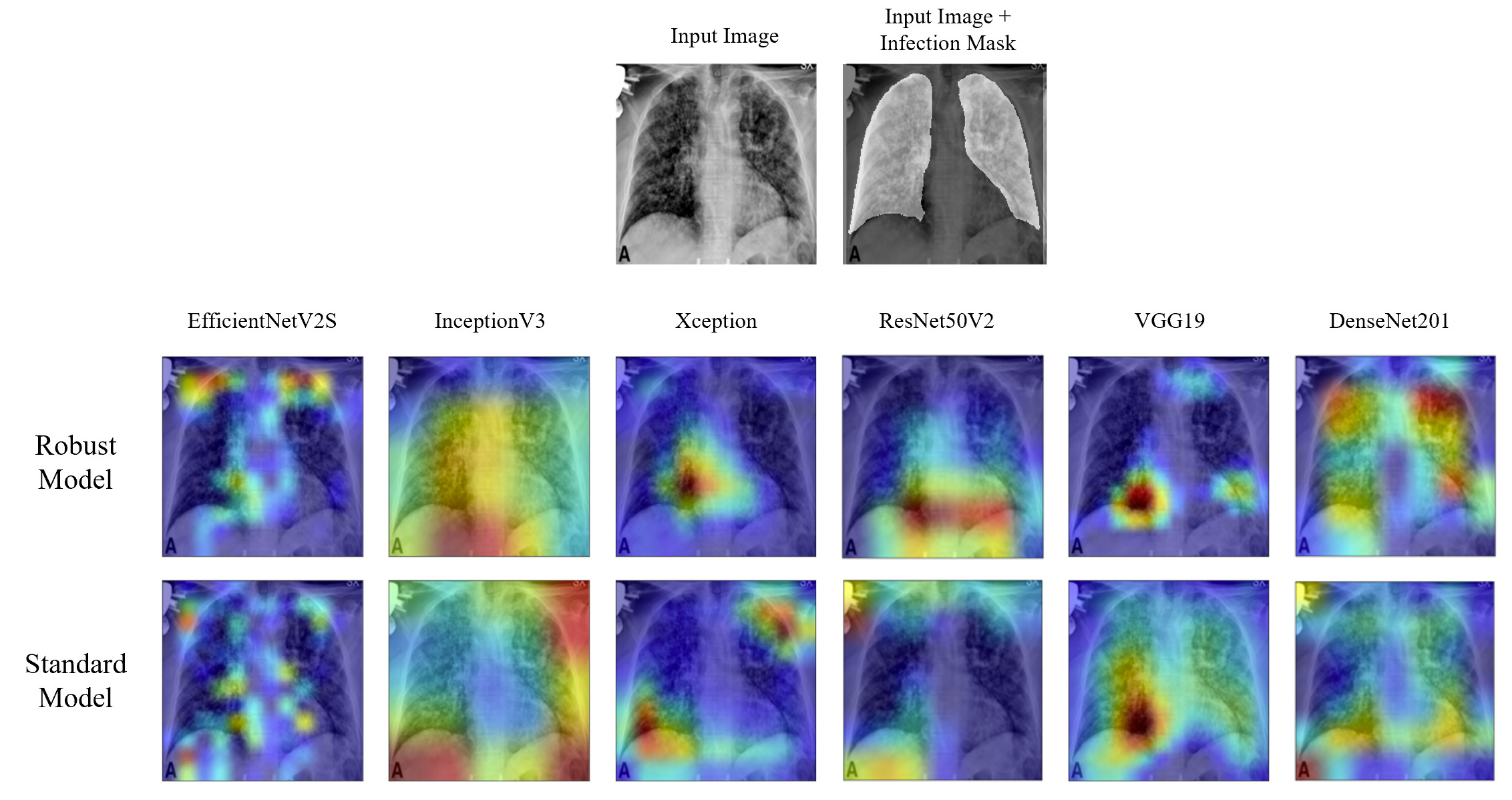}
\end{figure}

We make further comparisons between standard and robust Grad-CAM heatmaps of VGG19 due to their generally higher visual coherence on CXR images.

\subsubsection{Visual Coherence}

The robust VGG19 model yields sharper and more visually coherent Grad-CAM heatmaps. We notice that the robust heatmaps yield higher saliency areas that are more focused and concentrated in the lung region.

\begin{figure}[H]
\centering
\caption{Figure 3: Standard and Robust VGG19 Grad-CAMs: Visual Coherence}
\includegraphics[scale = 0.9]
{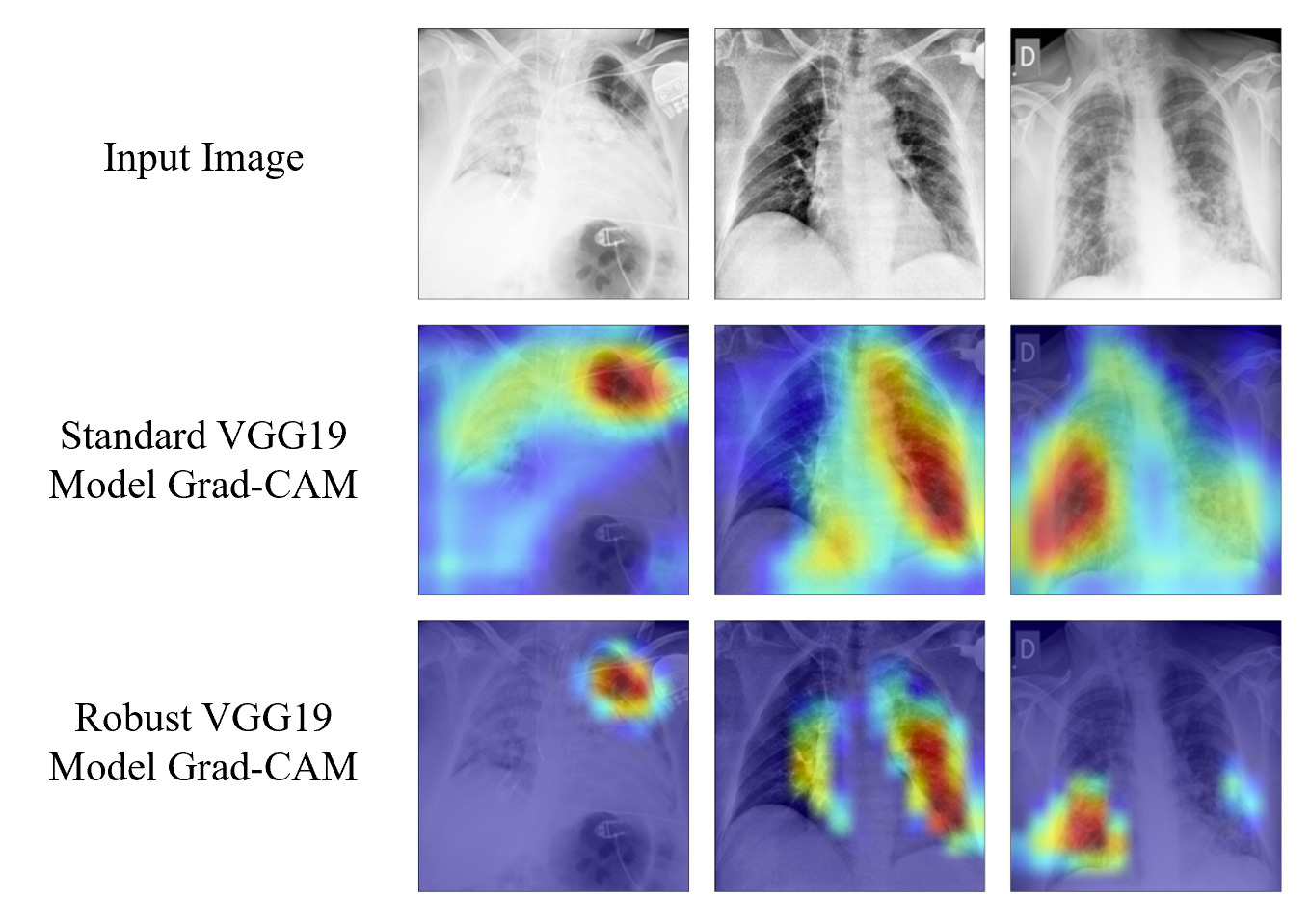}
\end{figure}

\subsubsection{External Text Annotations}

Adversarially trained models are more robust against external annotations on a CXR image (see top left of each CXR image in Figure 4). Though both the standard and robust model highlight the annotation region (indicating its positive correlation with the COVID-19 class), the robust VGG19 model attributes further importance to relevant regions within the lung. 

\begin{figure}[H]
\centering
\caption{Figure 4: Standard and Robust VGG19 Grad-CAMs: External Text Annotations}

\includegraphics[scale = 0.9]
{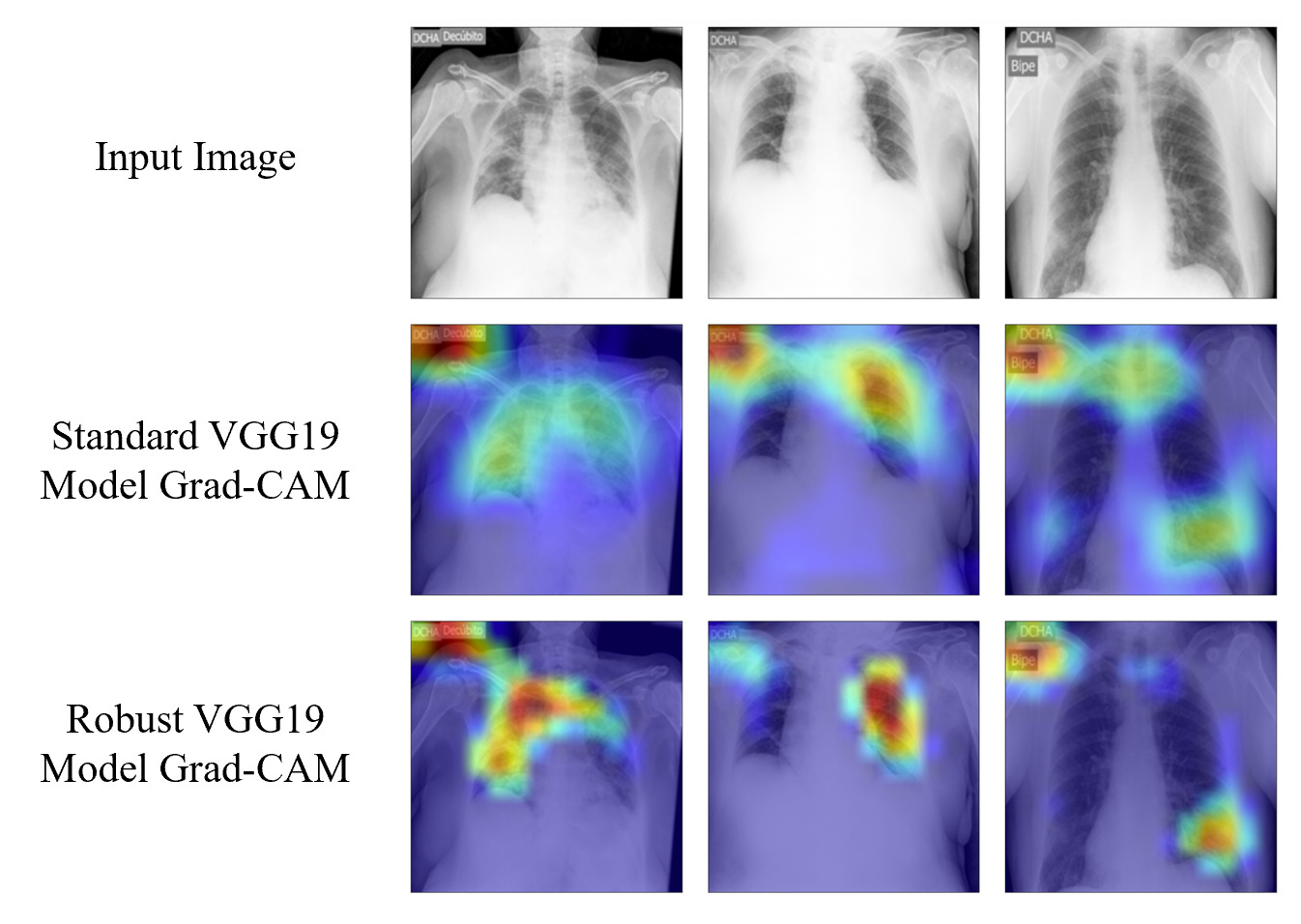}
\end{figure}

\subsubsection{Radiologist Annotations}

Robust models exhibit higher visual alignment with radiologist annotations. The standard and adversarially trained model both attribute high saliency values within regions denoted by infection masks (row 2 of Figure 5). Robust model heatmaps, however, contain areas of high saliency values that are more focused and visually agreeable with radiologist annotations.

\begin{figure}[H]
\centering
\caption{Figure 5: Standard and Robust VGG19 Grad-CAMs: Radiologist Annotations}
\includegraphics[scale = 0.9]{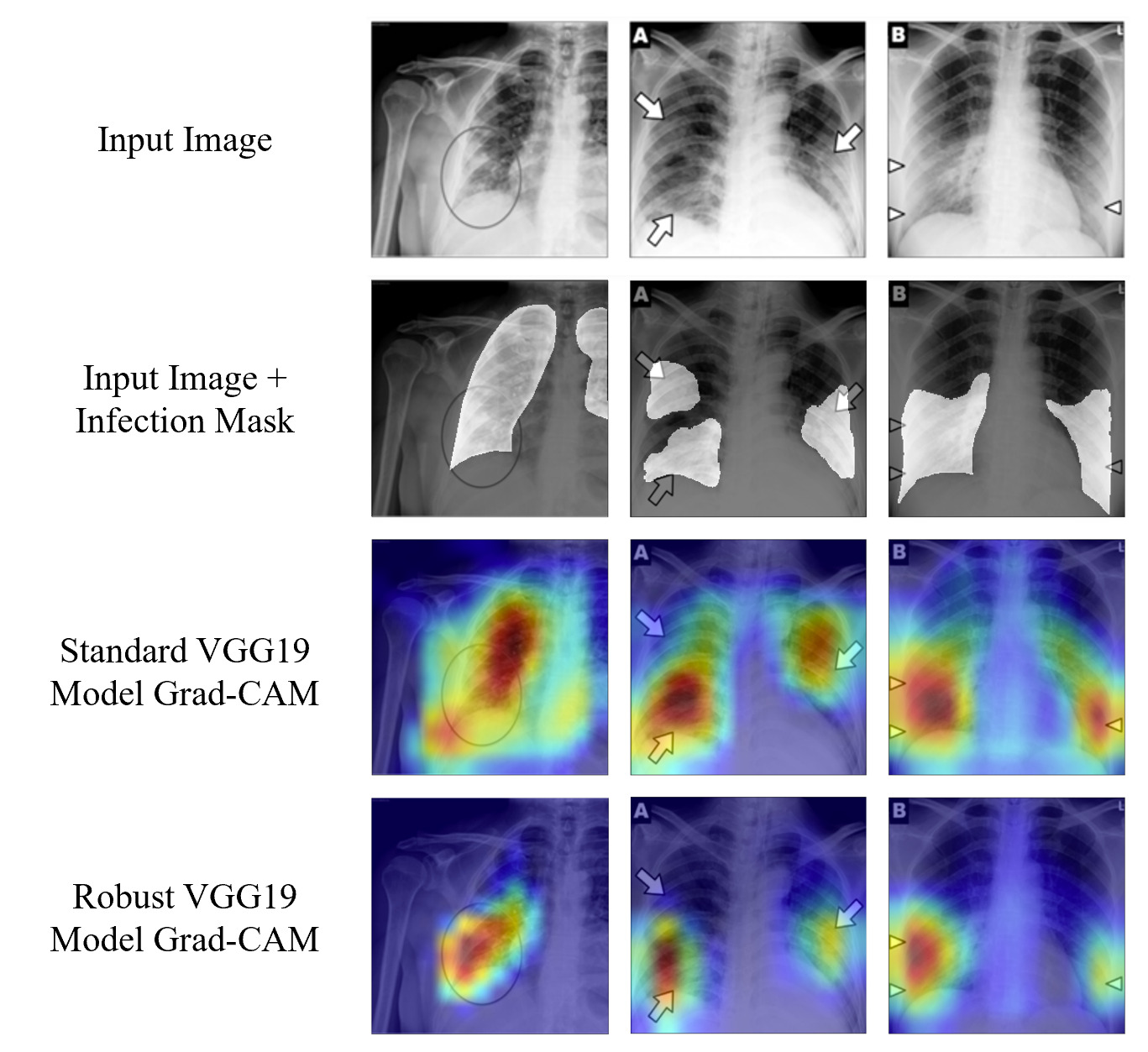}
\end{figure}

\section{Discussion}
\label{Discussion}
As we transition into an endemic phase of COVID-19, it is essential to continue to mitigate the risk of potential COVID-19 surges. The proposed methods ensure timely diagnoses of COVID-19, and can also be used to build more resilient healthcare systems for future public health threats.

In comparison to the current gold-standard RT-PCR test, which achieves a sensitivity of 0.71 when diagnosing COVID-19 \citep{fang2020sensitivity}, the proposed EfficientNetV2S model obtains up to 0.9703 accuracy, 0.9923 precision, 0.9797 recall, and an F1 score of 0.9860 across three classes. The increased sensitivity, or recall, minimizes the rate of false negatives during classification, which would be detrimental during the ongoing public health crises \citep{west2020covid}, since receiving immediate treatment for preventing secondary infections due to COVID-19 is critical for patient wellbeing.

We also note that Xception attains 99.95\% precision; out of the 6788 CXR images in the test set, the model only misclassifies 3 non-COVID-19 images into the COVID-19 class (on average). We hypothesize that its high predictive power may be due to its small 3x3 kernel sizes and the limited number of layers, which is suitable for extracting nuanced details in chest X-ray images. While the detrimental impact of false negatives are routinely discussed in regards to COVID-19, there are also consequences of false positives, including physical, financial, mental, and societal problems \citep{surkova2020false}. Given the common shortages of clinical supplies for treating COVID-19 during peaks of the pandemic, the high precision value of models such as Xception can maximize the effective distribution of hospital resources.

Due to the high-stake nature of accurate disease diagnosis and prognosis, it is essential that AI models are robust prior to clinical implementation. Standard deep learning models, especially for biomedical applications, are vulnerable to adversarial attacks, and often misclassify images that contain imperceivable pixel-level perturbations, which could result in detrimental consequences for patients \citep{ghaffari2022adversarial}. For instance, on perturbed images (with $\epsilon = 0.02$), the highest classification accuracy is a mere 68.95\% (achieved by the standard ResNet50V2 model). Higher intensity perturbations (greater $\epsilon$) on the images, albeit more noticable to the human eye, will only further decrease the classification accuracy. Evidently, this critical fault in standard deep learning models will cause high levels of misdiagnosis, leading to poor medical outcomes, patient mental distress, and potential legal consequences. Currently, there are few studies that consider adversarial attacks when developing machine learning models, leaving an unmet clinical need \citep{foote2021now,ghaffari2022adversarial}. The incorporation of adversarial training here provides major steps toward implementing machine learning models in diagnostic routines relating to COVID-19.

Importantly, we note that the robust EfficientNetV2S model mirrors the performance level of its standard version, attaining the highest value for three of the same metrics, highlighting that the incorporation of adversarial training does not negatively affect prediction even in the absence of perturbations. For robust models tested on perturbed images, EfficientNetV2S has the highest accuracy of 0.9503, Xception achieves the highest precision of 0.9923, and DenseNet201 attains the highest recall and F1 score of 0.9878 and 0.9772, respectively. From a public health standpoint, recall may be the most important metric when evaluating diagnostic models, to provide early treatment and mitigate the spread of disease. Therefore, an adversarial DenseNet201 model offers higher levels of reliability in aiding clinical diagnosis compared to standard models.

In order to address the common stigma surrounding the black-box nature of CNN-based algorithms, we also apply Grad-CAM heatmaps to the standard and robust models, while further comparing the extent to which adversarially trained models can improve visual coherence and explainability. To qualitatively evaluate our methods, we utilized ground-truth radiologist annotations to ensure the model classified images based on important features. We chose the VGG19 model to make further visual comparisons, despite its slightly lower classification performance, since it uses features that are more closely aligned to gold standard annotations. This may be an important consideration in a clinical settings for more trustworthiness in AI explainability. Between the standard and robust VGG19 model, we found that robust models have sharper and more visually coherent Grad-CAM heatmaps. 

In contrast to standard model heatmaps, robust model visualizations present high saliency areas that are more contained inside the lung and further focused on ground glass opacities commonly associated with COVID-19. This is consistent with previous findings, which demonstrated that robust models’ learned representations tend to better align with salient data characteristics and human perception \citep{margeloiu2020improving, tsipras2018robustness}. With high saliency values predominantly in the lung, robust model Grad-CAMs yield more specific and trustworthy explanations, which can better aid doctors in making clinical decisions.

Adversarially trained models are more robust against the presence of textual annotations. Some CXR images, especially in publicly available datasets, contain annotations that may be strongly associated with the COVID-19 class. This could potentially cause models to use this textual information as a “reason” for making the correct prediction. Robust models, while also addressing the importance of the annotation, recognize lung-relevant features with significantly more weight than standard models. This indicates that robust models may indeed learn to make predictions based on relevant features, significantly boosting their level of reliability as an explainability tool \citep{ilyas2019adversarial, margeloiu2020improving}.

Moreover, robust models produce heatmaps that are more closely aligned to radiologist annotations present on the CXR image. Upon overlaying infection masks on the CXR images, we find that both standard and robust model GradCAM saliency areas are mostly contained within the infection regions. Robust model heatmaps, however, show areas of high saliency values that agree considerably more with visual radiologist annotations than standard model heatmaps. From this, we hypothesize that robust model heatmaps may signify sharper sub-regions that are either more indicative of or seriously impacted by COVID-19, and should therefore be further studied. Considering the high degree of alignment between radiologist annotations and robust model heatmaps, we can generally expect robust model explanations to be both human-meaningful and faithful to the models’ decision-making process.

\section{Conclusion and Future Work}
\label{Conclusion and Future Work}
Given the recent expansion of COVID-19 cases across the world, it is important to identify a quick and effective method for diagnosis. Previous works have achieved impressive results with AI-based algorithms, though such methods falter in the context of generalizability and explainability, creating a bottleneck between algorithmic solutions and practical use in clinical settings. In this paper, we reported deep learning frameworks for detecting COVID-19 from CXR images with up to 97.03\% accuracy, which were trained and tested on a large dataset comprised of over 33,000 images from 10+ repositories to encourage generalization. To further close the gap between the model’s decision-making process with that of humans, we additionally implemented adversarial training to boost model classification robustness as well as visual coherence in Grad-CAM heatmaps. Our results demonstrate that visual explanations of robust models are sharper, focus on robust features, and are significantly more aligned with radiologist findings. 

However, we recognize that our proposed methods are not ready to be deployed in real-time environments. Although we aimed to construct generalizable models, it remains vitally important to assure the clinical validity of our proposed models, where validation on some private databases may be helpful. We note that robust model Grad-CAM heatmaps are more visually coherent than their standard counterparts, but we cannot quantitatively measure the quality of the visualizations compared to expert radiologist explanations. However, for short-term impact, we envision our methods aiding clinical practitioners in COVID-19 diagnosis, by providing more accurate explanations regarding black-box AI models' decision-making process.

For future work, we hope to extend and compare our findings to different imaging techniques (e.g. computed tomography), as well as incorporating multimodal or time-series data for producing better informed classifications. Further, our Grad-CAM heatmap comparisons use the robust VGG19 models trained with an adversarial step size $\epsilon$ = 0.02, though we hope to explore a larger range of $\epsilon$ values and different model architectures.

\section{Acknowledgements}
This work was supported by the National Science Foundation under Award Number 2027456 (COVID-ARC).

\section{Author Contributions}
\textbf{Karina Yang}: Conceptualization, Methodology, Software, Validation, Formal Analysis, Investigation, Data curation, Writing - Original Draft, Visualization.

\textbf{Alexis Bennet}: Writing - review \& editing, Resources, Data curation, Supervision.

\textbf{Dominique Duncan}: Writing - review \& editing, Resources, Data curation, Supervision, Project administration,
Funding acquisition.

\bibliography{sample}

\end{document}